\documentclass[uplatex]{article}
\usepackage[affil-it]{authblk}

\usepackage{amsmath,amssymb,color,dcolumn,bm,cite}
\usepackage[hypertex,bookmarks=true]{hyperref}
\usepackage{setspace,atbegshi,color}
\usepackage[dvipdfmx]{graphicx}
\usepackage{dcolumn}
\usepackage{bm}
\usepackage{braket}
\usepackage{comment}
\usepackage{here}
\usepackage{geometry}
\newcommand{\hami}{\hat{H}}
\newcommand{\br}[1]{\left(#1\right)}
\newcommand{\sbr}[1]{\left[#1\right]}

\newcommand{\cre}[2]{\hat{#1}_{#2}^{\dagger}}
\newcommand{\ani}[2]{\hat{#1}_{#2}}
\newcommand{\dout}{D_{\mathrm{out}}}
\newcommand{\pout}{P_{\mathrm{out}}}

\begin{document}

	\title{Numerical Large Deviation Analysis of Eigenstate Thermalization Hypothesis}
	
	\author{Toru Yoshizawa}
	\author{Eiki Iyoda}%
	\author{Takahiro Sagawa}
	\affil{%
		Department of Applied Physics, the University of Tokyo
	}%

	\date{\today}
	
	\maketitle
	
	\begin{abstract}
	A plausible mechanism of thermalization in isolated quantum systems is based on the strong version of the eigenstate thermalization hypothesis (ETH), which states that all the energy eigenstates in the microcanonical energy shell have thermal properties. We numerically investigate the ETH by focusing on the large deviation property, which directly evaluates the ratio of athermal energy eigenstates in the energy shell. As a consequence, we have systematically confirmed that the strong ETH is indeed true even for near-integrable systems. Furthermore, we found that the finite-size scaling of the ratio of athermal eigenstates is double exponential for non-integrable systems. Our result illuminates universal behavior of quantum chaos, and suggests that large deviation analysis would serve as a powerful method to investigate thermalization in the presence of the large finite-size effect.
	\end{abstract}

\maketitle

\textit{Introduction.}
The mechanism of thermalization has been a central issue in statistical physics, which bridges a gap between quantum mechanics and thermodynamics. This problem was first tackled by von Neumann in 1929 \cite{VonNeumann2010}, and has been extensively studied in recent years. Especially, it has been numerically shown that an isolated quantum system thermalizes when the system is non-integrable \cite{Rigol2008,Manmana2007,Moeckel2008,Eckstein2009}, while it does not thermalize when it is integrable \cite{Rigol2006,Rigol2007,Manmana2007,Rigol2009,Calabrese2011,Rigol2016,Vidmar2016}. The state-of-the-art technologies such as ultracold atoms \cite{Kinoshita2004,Paredes2004,Kinoshita2006,Trotzky2012,Kaufman2016} and superconducting qubits \cite{Neill2016} also provide an experimental platform to investigate dynamics of isolated quantum systems. 

A significant concept to understand the mechanism of thermalization is the strong eigenstate thermalization hypothesis (strong ETH) \cite{Rigol2008,Berry1977,Srednicki1994}, which states that all the energy eigenstates in the microcanonical (MC) shell are thermal. The word \textit{thermal} means that the expectation value of an observable equals the MC average: An energy eigenstate $\ket{E_{\alpha}}$ is thermal if it satisfies
\begin{equation}
\langle E_\alpha | \hat O | E_\alpha \rangle \simeq \langle \hat O \rangle_{\rm MC}.
\label{thermalization}
\end{equation}
We refer to an eigenstate as \textit{athermal} if it does not satisfy Eq.~(\ref{thermalization}).

We note that the long-time average of the expectation value coincides with the expectation value in the diagonal ensemble (DE) \cite{Tasaki1998,Reimann2008,Reimann2015}, where an initial state $\hat{\rho}_0=\sum_{\alpha,\beta}c_{\alpha}c_{\beta}^{\ast}\ket{E_{\alpha}}\bra{E_{\beta}}$ gives $\hat{\rho}_{\mathrm{DE}}=\sum_{\alpha}|c_{\alpha}|^2\ket{E_{\alpha}}\bra{E_{\alpha}}$ with $\ket{E_{\alpha}}$ being an energy eigenstate. Therefore, if all the eigenstates are thermal, the DE average of an operator $\hat{O}$ is equal to the MC average $\braket{\hat{O}}_{\mathrm{MC}}$:
\begin{equation}
{\rm tr}\sbr{\hat \rho_{\rm DE} \hat O}=\sum_{\alpha} |c_{\alpha} |^2 \braket{E_{\alpha} | \hat O | E_{\alpha} } \simeq\braket{\hat{O}}_{\mathrm{MC}},
\end{equation}
which guarantees that the long-time average equals the MC average. Thus, the strong ETH is a sufficient condition for thermalization. We note that, strictly speaking, we need to assume that the width of the energy distribution is at most subextensive in order to obtain Eq.~(2). Based on the various numerical results on the ETH, it is believed that the strong ETH is true for non-integrable systems \cite{Rigol2008,Steinigeweg2014,Beugeling2014,Kim2014,Eckstein2009,Mondaini2016,Steinigeweg2013,Khodja2015,Dymarsky2018}, but not true for integrable systems~\cite{Rigol2008,Ikeda2013}. 

Another important concept related to thermalization is quantum chaos \cite{Mehta,Stockmann}. In this context, it has been established that the distribution of energy level spacings is the Wigner-Dyson distribution if the system is non-integrable, while it is Poissonian if the system is integrable. It has been shown that near-integrable systems with very small integrability-breaking terms are also quantum chaotic \cite{Rabson2004, Santos2010}.

It is important to investigate the relationship between the strong ETH and quantum chaos \cite{Rigol2010,DAlessio2016,Borgonovi2016, Rigol2009}. However, the finite-size effect is significant for near-integrable systems, which would make numerical verification of the strong ETH quite hard. Therefore, it is highly desirable to investigate a systematic method to clearly judge the validity of the strong ETH within reasonable computational resources.

In this Letter, we numerically investigate the ETH in the spirit of large deviation analysis \cite{Touchette2009}, which directly evaluates the ratio of athermal energy eigenstates in the MC energy shell. In general, large deviation analysis is a statistical method to analyze asymptotic behavior of atypical events that are largely deviated from typical values. Since athermal eigenstates are atypical in terms of the ETH, large deviation analysis directly fits the investigation of the ETH, and will reveal its finite-size scaling even in the presence of the large finite-size effect.

By applying large deviation analysis along with the numerical exact diagonalization, we systematically investigated the validity of the strong ETH. Specifically, we performed numerical calculations of two kinds of spin models in one dimension: the XX model (integrable) and XXX model (with and without integrability-breaking terms). As a result, we clearly confirmed that the strong ETH holds only for the non-integrable cases. Our central observation is that the strong ETH is valid for near-integrable systems with very small integrability-breaking terms, which supports that the strong ETH is true even with an infinitesimal integrability breaking term~\cite{Rigol2009}. This illuminates the direct correspondence between the strong ETH and quantum chaos~\cite{Santos2010}.

Furthermore, we found that the ratio of athermal eigenstates shows a double exponential decay in the system size for non-integrable systems, which is contrastive to the exponential decay in integrable systems. In addition, we found that an exponent of the double exponential is almost independent of the degree of integrability breaking. We also argue that the double exponential behavior is qualitatively consistent with a typicality argument of the Hilbert space~\cite{Popescu2006,VonNeumann2010,Reimann2015}.

\textit{Large deviation analysis.}
We now formulate large deviation analysis in our setup. We consider a one-dimensional spin system on a lattice with $L$ sites. Let $\mathcal{M}(E, \Delta E)$ be the set of energy eigenstates in the energy shell with $\left[E-\Delta E , E\right]$, and $D := \left| \mathcal{M}(E, \Delta E)\right|$ be the dimension of the energy shell. We denote the expectation value with an eigenstate $\ket{E_{\alpha}}$ by $\braket{\hat{O}}_{\alpha}$ and the MC average by $\braket{\hat{O}}_{\mathrm{MC}}$. We fix a small positive constant $\varepsilon > 0$. We define that eigenstate $\ket{E_{\alpha}}$ is \textit{athermal} with respect to observable $\hat{O}$, if the difference between $\braket{\hat{O}}_{\alpha}$ and $\braket{\hat{O}}_{\mathrm{MC}}$ is larger than threshold $\varepsilon$, i.e., $\left| \braket{\hat{O}}_{\alpha} - \braket{\hat{O}}_{\mathrm{MC}}\right| > \varepsilon$. Then, the number of athermal eigenstates in the energy shell, denoted by $\dout$, is given by

\begin{equation}
\dout:=\sum_{\ket{E_{\alpha}}\in \mathcal{M}(E, \Delta E)}\theta\br{\left|\braket{\hat{O}}_{\alpha}-\braket{\hat{O}}_{\mathrm{MC}}\right|> \varepsilon},
\label{dout}
\end{equation}
where $\theta (\cdot )$ is the step function (i.e., $\theta (x) = 0$ ($x \leq 0$) and $\theta (x) =1$ ($x >0$)).

The strong ETH implies that $D_{\rm out}$ is zero if $L$ is sufficiently large.  On the other hand, there is a mathematical theorem \cite{Ogata2010,Tasaki2016,Mori2016}, often referred to as the \textit{weak} ETH \cite{Biroli2010,Iyoda2017,Ikeda2011,Alba2015}, which states that the ratio of athermal eigenstates decays at least exponentially in $L$:
\begin{equation}
\frac{\dout}{D}\le e^{-\gamma L},
\label{weak_ETH}
\end{equation}
where $\gamma > 0$ is a constant with $\gamma \propto \varepsilon^2$. We note that inequality (\ref{weak_ETH}) has been mathematically proved only for local observables \cite{Ogata2010,Tasaki2016,Mori2016}.

While inequality (\ref{weak_ETH}) gives an upper bound of $\dout/D$, the actual scaling of $D_{\rm out}/D$ is nontrivial. If $\dout/D$ decays exponentially, $\dout$ itself increases exponentially in $L$ and never goes to zero even in the thermodynamic limit, because $D$ increases exponentially in $L$ \cite{Toda}. On the other hand, if $\dout/D$ decays faster than exponentially, it is reasonable to consider that $\dout$ will become zero with large but finite $L$, because such fast decay much dominates the exponential increase of $D$. This is the case that the strong ETH is true, and therefore we can judge the validity of the strong ETH based on the finite-size scaling of $\dout/ D$.

\begin{figure}[t]
	\centering
	\includegraphics[width=0.5\linewidth]{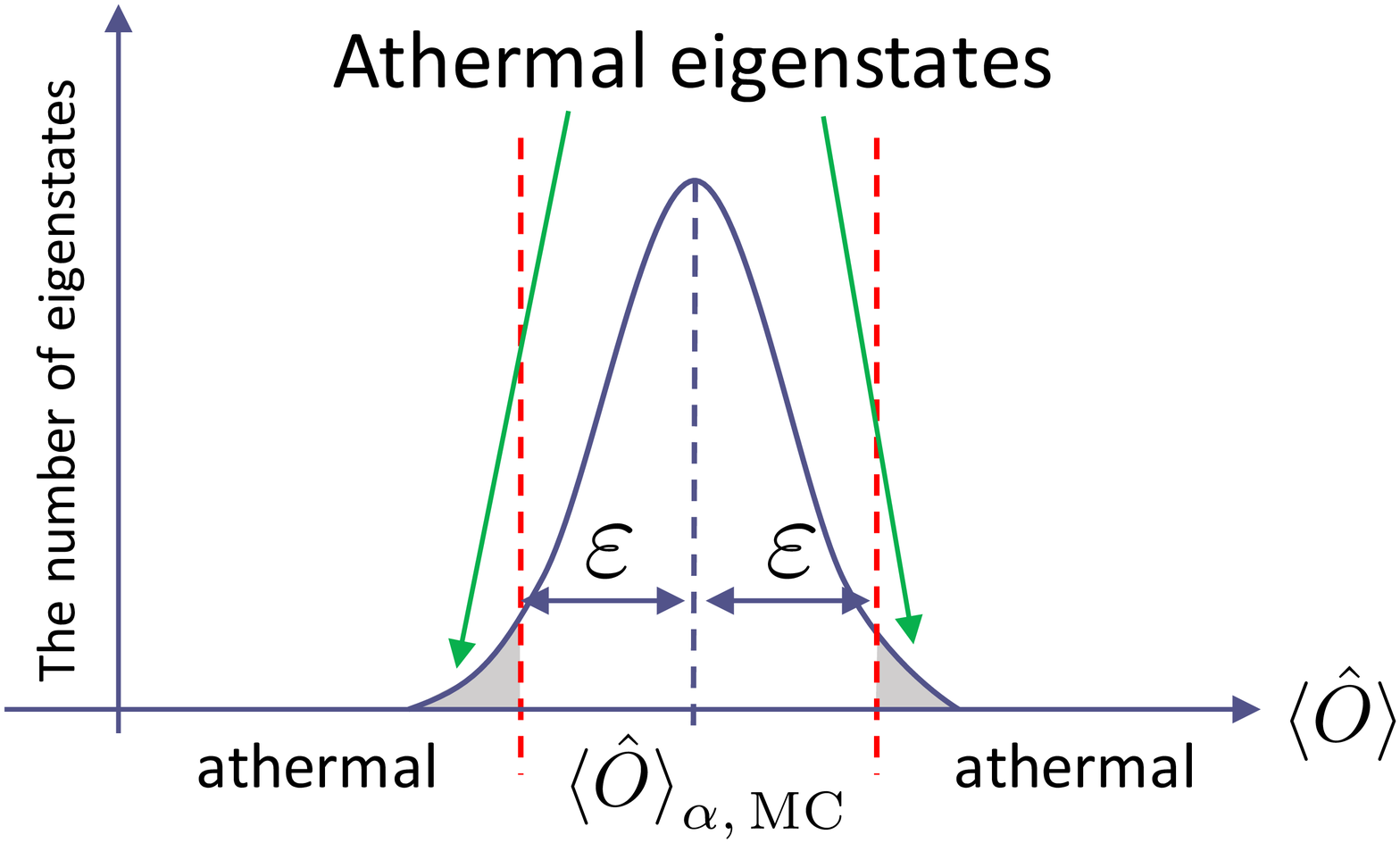}
	\caption{Schematic of our large deviation analysis. An energy eigenstate $\ket{E_{\alpha}}$ is athermal with respect to an observable $\hat{O}$, if the difference between $\braket{\hat{O}}_{\alpha}$ and $ \braket{\hat{O}}_{\alpha,\,\mathrm{MC}}$ is larger than a fixed threshold $\varepsilon$. We directly evaluate the ratio of athermal eigenstates, $\dout/D$, by numerical exact diagonalization.}
	\label{large_deviation}
\end{figure}

We here make a technical remark. If the width of the energy shell $\Delta E$ is too large, $\braket{\hat{O}}_\alpha$ changes largely with $\alpha$ within the energy shell. In such a case, we cannot precisely evaluate the finite-size scaling of $\dout/D$. On the other hand, if $\Delta E$ is too small, we cannot obtain the sufficient number of eigenstates for evaluating the scaling. To overcome this dilemma, we shift the MC average for each $\alpha$:
\begin{equation}
\braket{\hat{O}}_{\alpha,\,\mathrm{MC}}:=\frac{1}{|\mathcal{M}(E_{\alpha}, \delta)|}\sum_{\ket{E_{\beta}}\in\mathcal{M}(E_{\alpha}, \delta)}\braket{\hat{O}}_{\beta}, 
\end{equation}
where $\delta > 0$ is taken to be independent of $L$. Then, our modified definition of $\dout$ is given by 
\begin{equation}
\dout:=\sum_{\ket{E_{\alpha}}\in \mathcal{M}(E, \Delta E)}\theta \br{\left|\braket{\hat{O}}_{\alpha}-\braket{\hat{O}}_{\alpha,\,\mathrm{MC}}\right|-\varepsilon}.
\end{equation}
Such a definition is also adopted in Ref. \cite{Beugeling2014} to discuss the finite-size scaling of the variance of $\braket{\hat O}_\alpha$.

\textit{Model and method.}
To demonstrate large deviation analysis, we consider two kinds of one-dimensional spin chains. The first model is the XX model, which is integrable and is equivalent to free hard-core bosons (HCB). Let $\hat{b}_i$ and $\hat{b}_i^{\dagger}$ be the annihilation and the creation operators of a HCB at site $i$ ($=1,2, \cdots, L$). Their commutation relations are given by $[\cre{b}{i},\ani{b}{j}]=[\cre{b}{i},\cre{b}{j}]=[\ani{b}{i},\ani{b}{j}]=0$ $(i\ne j)
$, and $\{\cre{b}{i},\ani{b}{i}\}=1,\{\cre{b}{i},\cre{b}{i}\}=\{\ani{b}{i},\ani{b}{i}\}=0
$. The XX Hamiltonian is given by 
\begin{equation}
\hami_{XX}:=-\sum_{i=1}^{L}\sbr{\cre{b}{i}\ani{b}{i+1}+h.c.},
\label{XX_hamiltonian}
\end{equation}
where we adopted the periodic boundary condition $\ani{b}{L+1}=\ani{b}{1}$. This Hamiltonian can be transformed into the quadratic form of free fermions by the Jordan-Wigner transformation, which enables us to calculate large systems by using the Slater determinant \cite{Rigol2004}. In this model, we fix the filling of HCB at 1/8. 

The second model is the XXX model with and without next-nearest neighbor interactions. Let $\hat{n}_i := \hat{b}_i^{\dagger} \hat{b}_i$ be the HCB number operator at site $i$. The Hamiltonian is given by
\begin{align}
&\hami_{XXX}:=\frac{1}{1+\lambda}\sbr{\hami_{0}+\lambda\hat{W} },\label{XXZ_hamiltonian}\\
\hami_{0}&:=-\sum_{i=1}^{L}\sbr{\cre{b}{i}\ani{b}{i+1}+h.c.} +\sum_{i=1}^{L}\hat{n}_i\hat{n}_{i+1},\nonumber\\
\hat{W}&:=-\sum_{i=1}^{L}\sbr{\cre{b}{i}\ani{b}{i+2}+h.c.} +\sum_{i=1}^{L}\hat{n}_i\hat{n}_{i+2},\nonumber
\end{align}
where we again adopted the periodic boundary condition $\ani{b}{L+1}=\ani{b}{1}$. Here, $\hami_0$ is the original XXX Hamiltonian, which is integrable by the Bethe ansatz. $\hat{W}$ represents next-nearest interactions, which break the integrability of the system; $\lambda >0$ represents the degree of integrability breaking. The filling is fixed at 1/3 in this model. We note that the energy-level spacing distribution is well fitted by the Wigner-Dyson distribution even at small $\lambda$, like $\lambda=0.06$, when $L=24$ \cite{Santos2010}. 

Because of the translation invariance, we can block-diagonalize the Hamiltonian (\ref{XXZ_hamiltonian}) to $L$ sectors, and we performed numerical diagonalization within individual sectors. Because of the restriction of numerical resources, we used the following two different diagonalization methods. For $L=12$ to $21$ (for all $\lambda$) and $L=24$ ($\lambda=1$), we used the exact full diagonalization method. For $L=24$ with $\lambda \ne 1$, we used the Sakurai-Sugiura (SS) method \cite{Sakurai2003,Nagai2013}, which is a kind of the Krylov subspace method. By the SS method, we numerically calculated the energy eigenstates only within the target energy shell. 

We next discuss our choice of observables, for which we test the strong ETH. We note that different observables give different athermal eigenstates in general. We focus on the following three observables with different degrees of locality. The first is the neighboring correlation of the number operators $\hat{n}_1\hat{n}_2:=\cre{b}{1}\ani{b}{1}\cre{b}{2}\ani{b}{2}$, which is the most local observable over the three. The second is the third-nearest neighbor hopping $\hat{h}:=\frac{1}{L}\sum_{i=1}^{L}\sbr{\cre{b}{i}\ani{b}{i+3}+h.c.}$, which is the sum of slightly more non-local observables. The third is the zero-wavenumber momentum distribution $\hat{f}_0=\frac{1}{L}\sum_{i,j}\cre{b}{i}\ani{b}{j}$, which is non-local.

\begin{figure}[t]
	\centering
	\includegraphics[width=0.5\linewidth]{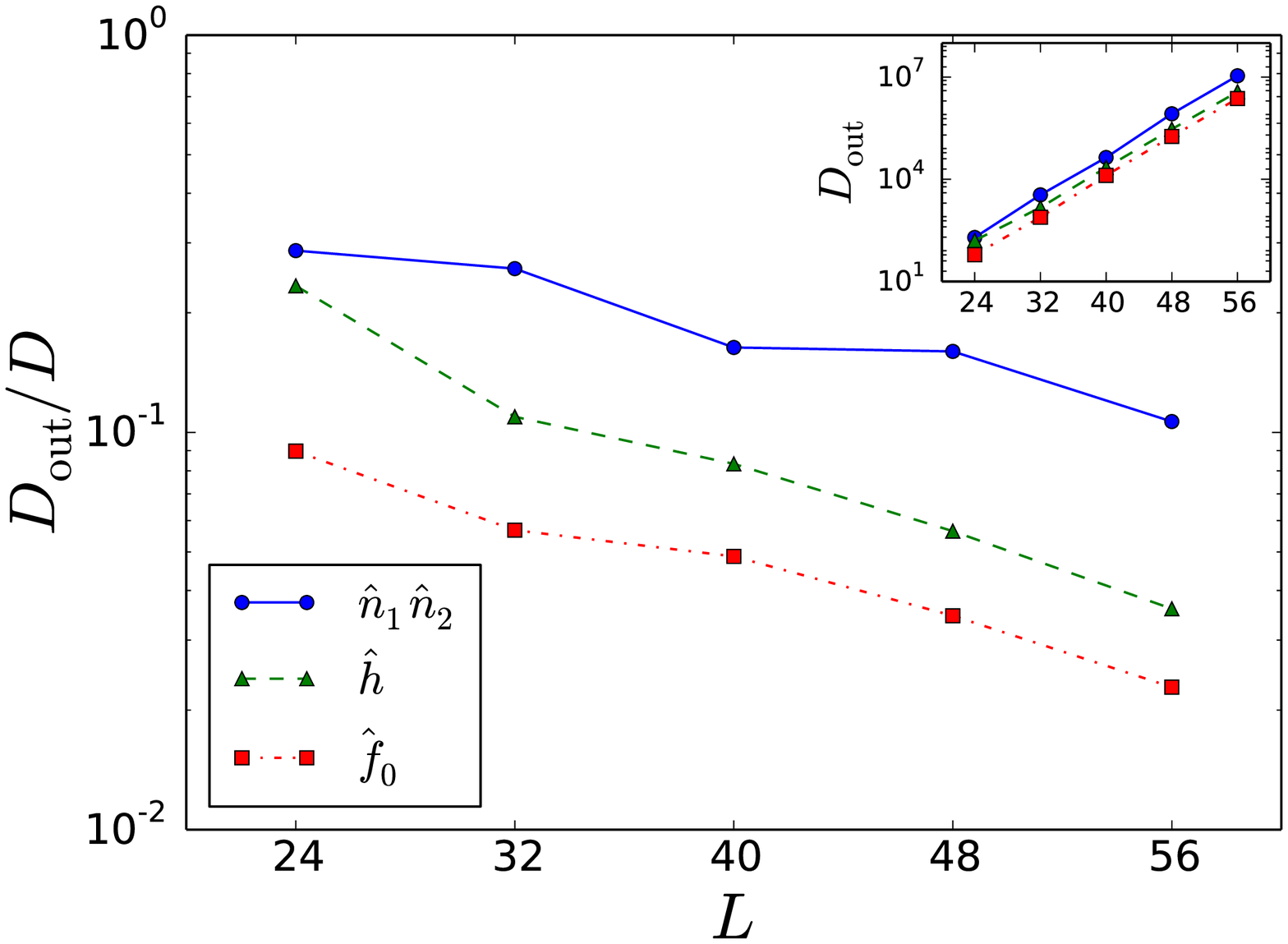}
	\caption{The system-size depencence of $\dout/D$ in the integrable system (\ref{XX_hamiltonian}), which decays exponentially. The inset shows the exponential growth of $\dout$. The parameters are set to $E=0,$ $\Delta E = 0.1L,$ $\delta=0.1$, and $\varepsilon=0.01,\,0.05, 0.2$ for $\hat{n}_1\hat{n}_2$, $\hat{h}$, $\hat{f}_0$, respectively.}
	\label{result_XX}
\end{figure}
\textit{Result.}
We first show our results of the integrable XX model (\ref{XX_hamiltonian}). Figure \ref{result_XX} shows the $L$-dependence of $\dout/D$ for the XX model. We observed that $\dout/D$ tends to decay exponentially, while a large amount of fluctuations remains. The flucuations would be caused by the large finite-size effect due to the existence of local conserved quantities. In addition, the $L$-dependence of $\dout$ itself is shown in the inset of Fig.~\ref{result_XX}. We clearly see that $\dout$ increases exponentially, and therefore the strong ETH is not true for all of the three observables. We emphasize that this exponential scaling of $D_{\rm out}/D$ is not mathematically trivial, because the weak ETH (\ref{weak_ETH}) only gives an upper bound.

We note that the absence of thermalization in integrable systems \cite{Rigol2007} stems from the fact that athermal eigenstates have large weights after quantum quench.  Since the ratio of athermal eigenstates is exponentially small, sampling of energy eigenstates by quench is exponentially biased \cite{Biroli2010} (see Supplemental Material for our numerical results). The origin of this behavior is the existence of the extensive number of local
conserved quantities \cite{Vidmar2016,Pereira2014}, leading to the generalized Gibbs ensemble \cite{Rigol2007}.

\begin{figure}[t]
	\centering
	\includegraphics[width=0.5\linewidth]{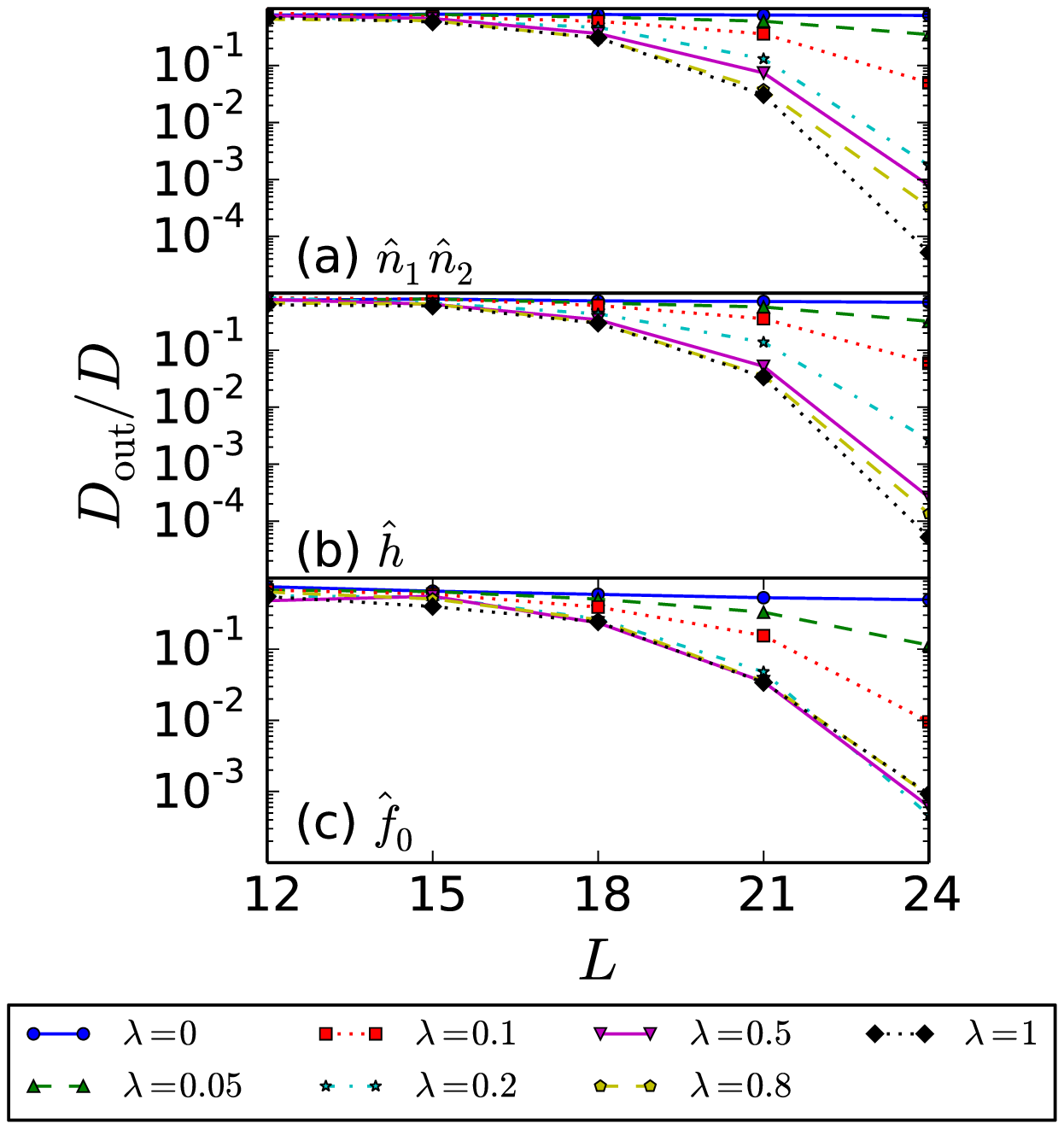}
	\caption{The system-size dependence of $\dout/D$ for the XXX model (\ref{XXZ_hamiltonian}) for several $\lambda$. The observables are given by (a)~$\hat{n}_1\hat{n}_2$, (b)~$\hat{h}$, and (c)~$\hat{f}_0$. The parameters are set to $E=0$, $\Delta E=0.1L$, $\delta=0.1$, and $\varepsilon=0.003,\,0.015, 0.05$ for (a), (b), (c), respectively.}
	\label{result_XXZ}
\end{figure}

We next show our results of the XXX model (\ref{XXZ_hamiltonian}). Figure \ref{result_XXZ} (a)-(c) show the $L$-dependence of $\dout/D$ for the three observables with several values of $\lambda$. We clearly see that the curves are concave for all the cases except for $\lambda = 0$, which implies that $\dout/D$ decays faster than exponentially if the system is non-integrable. As discussed before, it is reasonable to conclude that the faster-than-exponential decay of $\dout/D$ implies the strong ETH: $\dout = 0$ for sufficiently large $L$. More strictly, we can at least definitely exclude the possibility of the exponentially many athermal eigenstates (i.e., $\dout = e^{O(L)}$), which makes a sharp contrast to integrable systems.

\begin{figure}[t]
	\centering
	\includegraphics[width=0.5\linewidth]{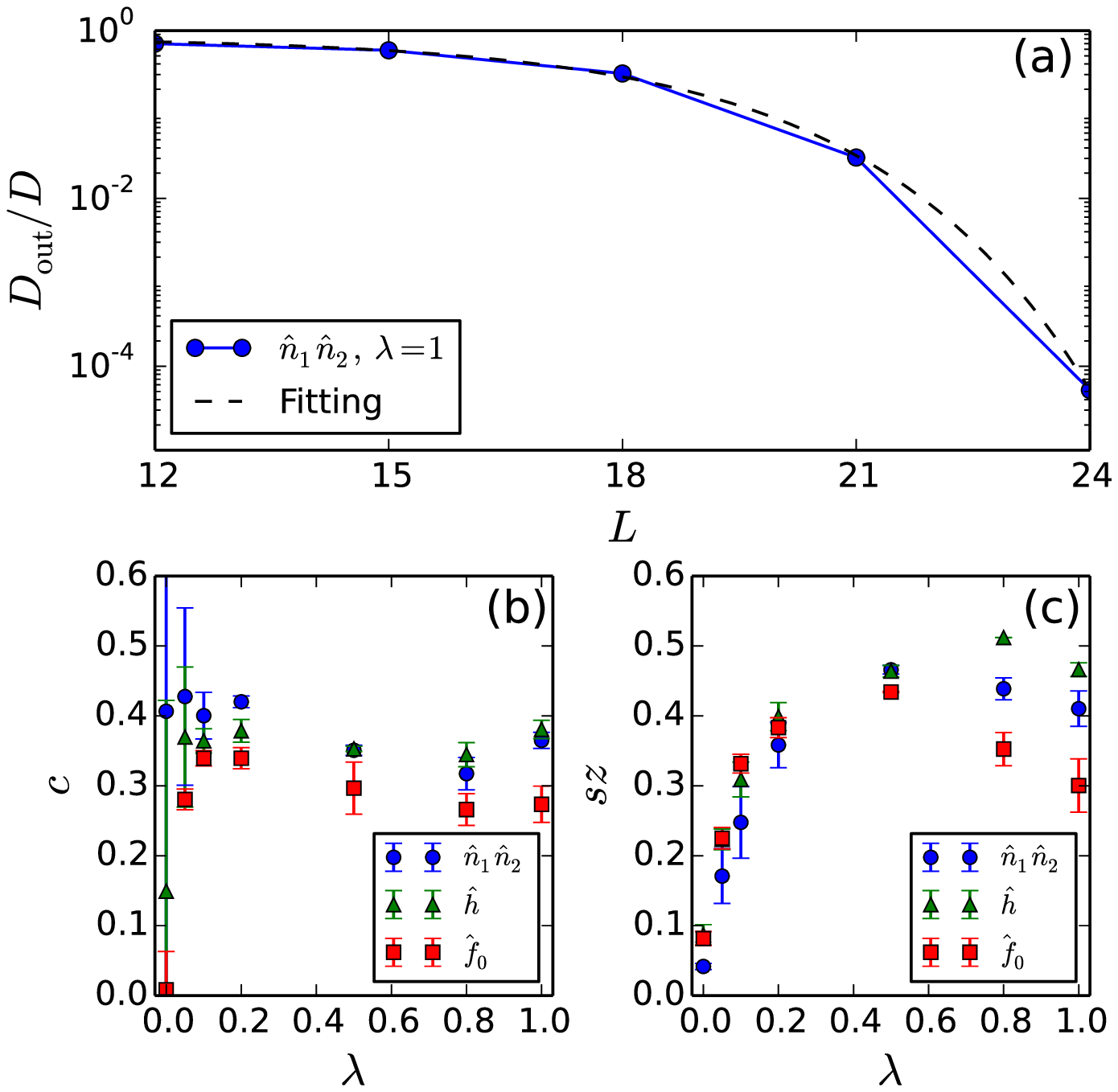}
	\caption{Numerical results of the fitting. (a)~An example of the fitting for $\hat{n}_1\hat{n}_2$ and $\lambda=1$. (b)~The $\lambda$-dependence of exponent $c$. We observe that $c>0$ for any $\lambda>0$, while the fitting fails for $\lambda = 0$ (integrable case). (c)~The $\lambda$-dependence of $sz$.   }
	\label{fitting}
\end{figure}

Interestingly, $\dout/D$ seems to decay double exponentially if $\lambda>0$. To quantitatively check this, we performed the fitting of $\log\sbr{D_{\mathrm{out}}/D}$ with a fitting function of the form $f(L)=a-be^{cL}$. Figure~\ref{fitting}~(a) shows the fitting for the case of observable $\hat{n}_1\hat{n}_2$ and $\lambda=1$, which seems to work well, while we have three fitting parameters for five data points. Because we can approximate $e^{-be^{cL}}\simeq e^{-b'L}$ for $c\ll1$ up to a prefactor, the parameter $c$ represents non-integrability of the Hamiltonian. We show the $\lambda$-dependence of $c$ in Fig.~\ref{fitting}~(b). The error bar shows the asymptotic standard error of the fitting, which is small if the fitting works well. The fitting fails at $\lambda=0$, as expected. On the other hand, $c$ has a positive value for any $\lambda \neq 0$, which suggests that the exponent $c$ becomes positive for an infinitesimal $\lambda > 0$. We can also see that the $\lambda$-dependence of $c$ is small especially for $\lambda \geq 0.5$. Surprisingly, $c$ takes almost the same values for $\hat n_1 \hat n_2$ and $\hat{h}$, while $c$ takes slightly different value for $\hat{f}_0$. These results suggest that $c$ takes a constant value independent of $\lambda > 0$ at least for each observable.

To consider the physical origin of the double exponential decay of $\dout /D$, we refer to the typicality argument of the Hilbert space \cite{Popescu2006,VonNeumann2010,Reimann2015}, from which $\dout/D$ is expected to decrease as $e^{-bD}$. We note that $D$ increases exponentially as $D=e^{sL}$, where $s$ is the entropy density. If this was the case, $c$ would equal $s$ that is independent of $\lambda$ and observables. However, the numerically obtained value of $c$ does not equal $s$, because $s\simeq0.6$ in our setup (see Supplemental Material for details). Therefore, the above naive typicality argument fails to quantitatively account for the double exponential behavior.

To further clarify the origin of this exponential behavior, we also evaluated the variance of $\langle \hat O \rangle_\alpha$, which is defined by~\cite{Beugeling2014}
\begin{equation}
\sigma_{\hat{O}}^2:= \frac{1}{D} \sum_{\ket{E_{\alpha}}\in \mathcal{M}(E, \Delta E)}\br{\braket{\hat{O}}_{\alpha}-\braket{\hat{O}}_{\mathrm{\alpha,\,MC}}}^2.
\label{variance}
\end{equation}
We calculated its $D$-dependence by fitting it with $\sigma^2_{\hat O} \propto D^{-z}$, and found that $z$ is smaller than unity.

By referring a general theory of large deviation~\cite{Touchette2009}, if the variance scales as $\sigma^2_{\hat O} \propto D^{-z}$, the large deviation shows the double exponential behavior of $D_{\rm out}/D \propto e^{-\gamma^\prime D^z} \simeq e^{-\gamma e^{szL}}$, implying $c=sz$. Correspondingly, the $\lambda$-dependence of $sz$ in our numerical result is shown in Fig.~\ref{fitting}~(c), which is close to $c$ for $\lambda > 0.5$. The deviation between $sz$ and $c$ for smaller $\lambda$ would be the finite-size effect.

On the other hand, if the above-mentioned typicality argument applies, $z=1$ should hold \cite{Sugita,Hams}. Our result $z<1$ thus implies that our system is not completely randomized in the entire Hilbert space.  This would be a consequence of the fact that the Hamiltonian and the observables are not random \textit{\`{a} la} Haar, but are connected with each other through physical structures such as locality.  In other words, we can regard $\tilde D:=D^{z}$ as an effective dimension of the Hilbert space in which our system is randomized.
To support this argument, we numerically calculated the operator norm of the commutator $\Delta_{\hat O} := \|i[\hami_{XXX}, \hat O]\|/\|\hat O\|$ for $\hat O = \hat n_1\hat n_2$, $\hat h$, and $\hat f_0$ (see Supplemental Material). The amount of $\Delta_{\hat O}$ exhibits a trend similar to Fig.~\ref{fitting}~(b) and (c). Specifically, in the strongly chaotic regime $\lambda \geq 0.5$, the order of the values of $\Delta_{\hat n_1\hat n_2}$, $\Delta_{\hat h}$, $\Delta_{\hat f_0}$ is the same as that of $c$ and $sz$. Therefore, we argue that a smaller value of $\Delta_{\hat O}$ implies a closer connection of $\hat O$ to the Hamiltonian $\hat H_{XXX}$, which reduces the effective randomness of $\hat O$ with respect to $\hat H_{XXX}$.

The foregoing results suggest that the strong ETH is true even with infinitesimal non-integrability. This is consistent with the level statistics in near-integrable systems \cite{Santos2010}. Our results directly support that a quantum chaotic system does thermalize even near integrability \cite{Rigol2009}. We note that our conclusions are valid independently of the choice of the parameters $E$, $\Delta E$, $\delta$ and $\varepsilon$ (see Supplemental Material). 

\textit{Conclusion.}
In this Letter, we have investigated large deviation analysis of the ETH, which enables us to directly evaluate the finite-size scaling of the ratio of athermal eigenstates. As a result, we found that the ratio of athermal eigenstates decays exponentially in the system size for the integrable case, while double-exponentially for the non-integrable case. The latter implies the strong ETH. In particular, we confirmed the validity of the strong ETH near integrable. We expect that large deviation analysis would further reveal fundamental properties of quantum chaos for a broader class of quantum many-body systems.

The authors are grateful to Marcos Rigol, Hal Tasaki and Youhei Yamaji for valuable discussions. E. I. is  grateful to Yasushi Shinohara for helpful comments on numerical calculation. The computation in this work has been done using the facilities of the Supercomputer Center, the Institute for Solid State Physics, The University of Tokyo. This work is supported by JSPS KAKENHI Grant No. JP16H02211, No. JP25103003, No. JP15K20944.

	\def\theequation{S\arabic{equation}}
	\def\thefigure{S\arabic{figure}}
	\setcounter{figure}{0}
	\setcounter{equation}{0}

	\clearpage
	\newpage
	\setcounter{page}{1}
	\begin{center}
		\vspace{10mm}
		\Large{Supplemental Material : Numerical Large Deviation Analysis of Eigenstate Thermalization Hypothesis}
		
		\vspace{5mm}
		\large{Toru Yoshizawa, Eiki Iyoda, and Takahiro Sagawa}
		
		\vspace{5mm}
		\large{\textit{Department of Applied Physics, the University of Tokyo}}
		
		\vspace{5mm}
	\end{center}
	
	In this Supplemental Material, we provide supplemental numerical results. In Sec. 1, we evaluate the sum of the weights of energy eigenstates after quantum quench for the integrable system (6) of the main text. As a result, we found that the weight distribution after the quench is exponentially biased, leading to the absence of thermalization in the integrable system. In Sec. 2, we show the dependence of $\dout/D$ on four parameters: $\varepsilon$, $E$, $\delta$, and $\Delta E$, in order to confirm that our conclusion is valid independently of the choice of these parameters, except for some extreme cases. In Sec~3, we show the $\lambda$-dependence of the norm of the commutator $\Delta_{\hat O}$ that quantifies a ``distance'' between the Hamiltonian $\hami_{XXX}$ and the observable $\hat O$. In Sec. 4, we show the values of entropy density $s$ in detail.
	
	\section{Large deviation analysis of the weights of energy eigenstates after quantum quench}
\begin{figure}[t]
		\centering
		\includegraphics[width=0.5\linewidth]{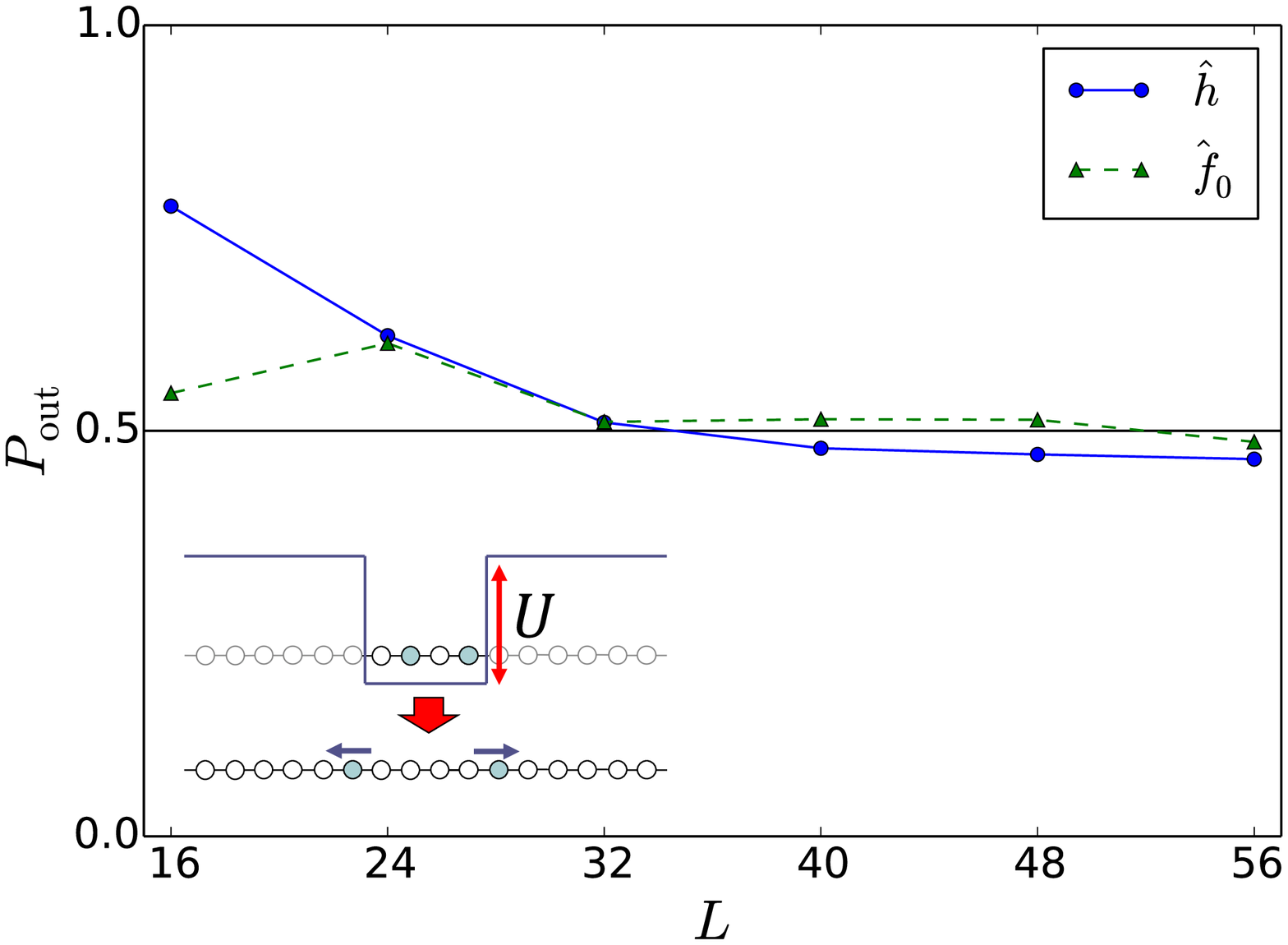}
		\caption{The system size-dependence of $\pout$ defined in Eq. (S1). (Inset) Schematic of the quantum quench: An infinitely high potential wall is removed.}
		\label{pout}
	\end{figure}

	We here consider the origin of the absence of thermalization after quantum quench in an integrable system [7]. In the absence of thermalization, an athermal eigenstate must have an exponentially larger weight than a typical weight of thermal eigenstates, because the number of athermal eigenstates is exponentially smaller than that of thermal eigenstates (i.e., $\dout/D$ decays exponentially). We can directly see this by large deviation analysis. Specifically, we consider the XX model (6) of the main text, and evaluate the sum of the weights of athermal eigenstates: 
	\begin{equation}
	P_{\mathrm{out}}:=\sum_{\alpha}\left|\braket{E_{\alpha}|\psi}\right|^2\theta \br{\left|\braket{\hat{O}}_{\alpha}-\braket{\hat{O}}_{\mathrm{MC}}\right|-\varepsilon},
	\end{equation}
	where $\ket{\psi}$ is the initial state before quench, and the summation is taken over all the eigenstates. As the initial state, we adopt the ground state of the Hamiltonian before the quench:
	\begin{equation}
	\hami_{0}:=-\sum_{i=1}^{L}\sbr{\cre{b}{i}\ani{b}{i+1}+h.c.}+U\sum_{i=1}^{3L/4}\cre{b}{i}\ani{b}{i},
	\end{equation}
	where we take the limit $U\rightarrow+\infty$. The inset of Fig. \ref{pout} shows a schematic of the quench.
	
	Figure~\ref{pout} shows our numerical result of $\pout$ for observables $\hat{h}$ and $\hat{f}_0$ defined in the main text, where $\varepsilon$ is taken to be $\varepsilon=\left|\mathrm{Tr}\sbr{\hat{O}\rho_{\mathrm{DE}}}-\braket{\hat{{O}}}_{\mathrm{MC}}\right|$. We do not show the result for $\hat{n}_1\hat{n}_2$ because the difference between the diagonal average and the MC average is too small to see the large deviation property of $\pout$. As seen from Fig.~\ref{pout}, $\pout$ approximately equals $1/2$ regardless of the system size for $L \geq 32$, which implies the absence of thermalization in the thermodynamic limit. We can thus conclude that the sampling of energy eigenstates by the quench is exponentially biased to athermal eigenstates.
	
	\newpage
	\section{The parameter-dependence of $\dout/D$}
	In this section, we show the parameter-dependence of the ratio of athermal eigenstates $\dout/D$ in order to check the validity of our conclusions in the main text. When we change one parameter to investigate the dependence of $\dout/D$ on it, we fix the other parameters to the same value as used in the main text. In the results shown below, the lack of the data point means that $\dout/D$ is exactly zero.
	
	\subsection{Integrable system}
	We first show the parameter-dependence of $\dout/D$ for the integrable system (6) of the main text. The parameters used in the main text are given by $E=0$, $\Delta E=0.1L$, $\delta=0.1$, and $\varepsilon=0.01, 0.05, 0.2$ for $\hat{n}_1\hat{n}_2$, $\hat{h}$, and $\hat{f}_0$, respectively.
	
\begin{figure}[H]
		\centering
		\includegraphics[width=\linewidth]{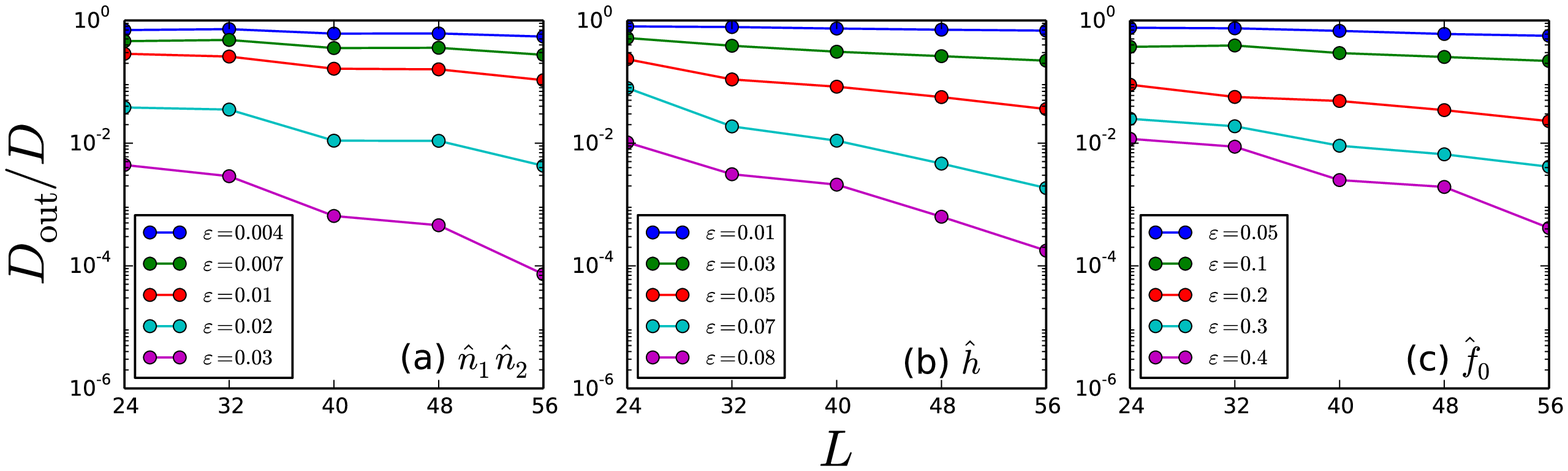}
		\caption{The $\varepsilon$-dependence of $\dout/D$ for the integrable system.}
		\label{int_epsilon}
	\end{figure}

	Figure \ref{int_epsilon} shows the $\varepsilon$-dependence of $\dout/D$, where $\dout/D$ decays exponentially for all $\varepsilon$. In addition, we found that $\dout/D$ with large $\varepsilon$ decays faster than with small $\varepsilon$, but never goes to zero. This implies that the weak ETH holds but the strong ETH does not hold.
	
\begin{figure}[H]
		\centering
		\includegraphics[width=\linewidth]{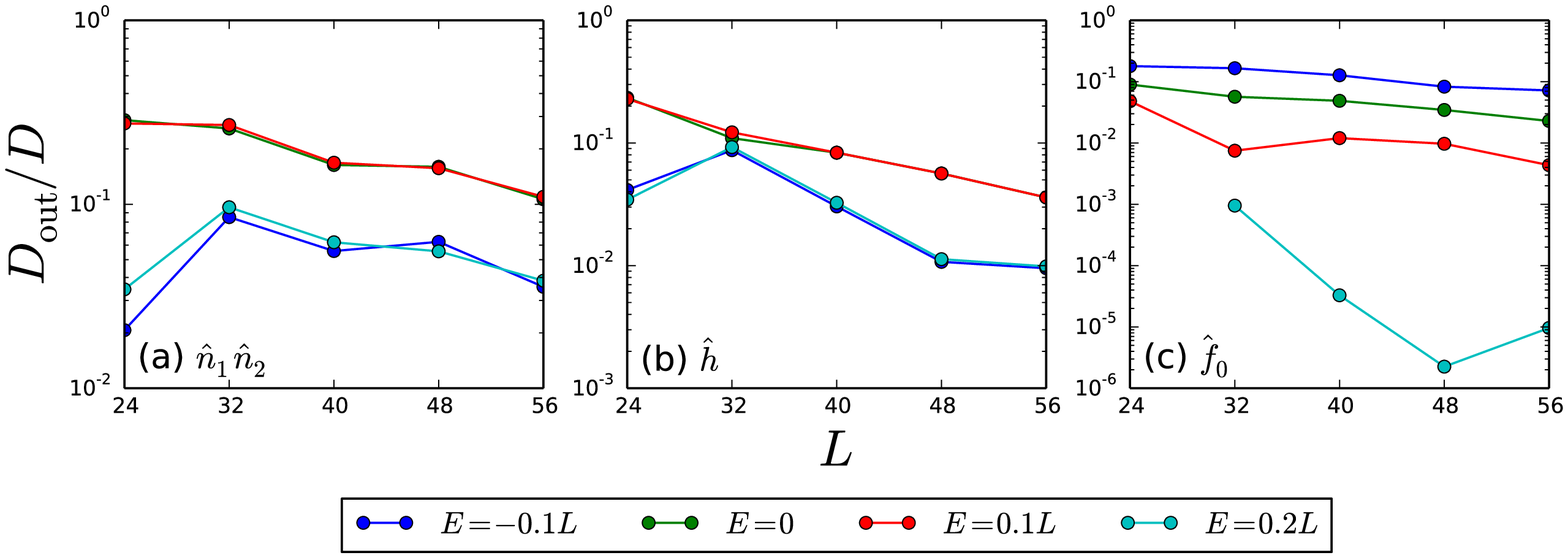}
		\caption{The $E$-dependence of $\dout/D$ for the integrable system.}
		\label{int_energy}
	\end{figure}

	Figure \ref{int_energy} shows the $E$-dependence of $\dout/D$, where $\dout/D$ decays exponentially except for the case of $\hat{f}_0$ with $E=0.2L$. A possible origin of this exceptional behavior is simply the large finite size effect; For a larger system, $\dout/D$ is expected to decay exponentially in any energy range. 
	
\begin{figure}[H]
		\centering
		\includegraphics[width=\linewidth]{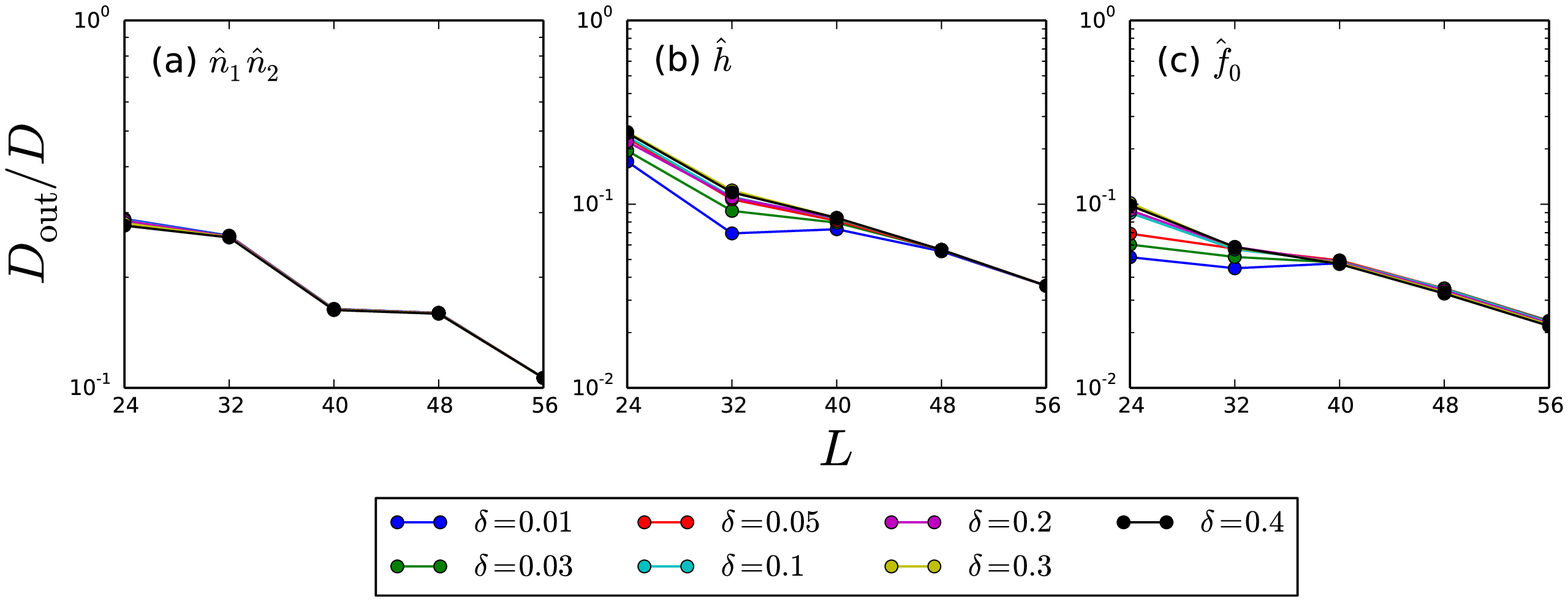}
		\caption{The $\delta$-dependence of $\dout/D$ for the integrable system.}
		\label{int_delta}
	\end{figure}

	Figure \ref{int_delta} shows the $\delta$-dependence of $\dout/D$. In the case that $\delta$ is too small, like $\delta=0.01$, the analysis does not work well for small $L$. This is because the number of energy eigenstates in the MC average is very small when $\delta$ and $L$ are too small. Even with such a small $\delta$, $\dout/D$ decays exponentially independent of $\delta$, when $L$ becomes large.
	
\begin{figure}[H]
		\centering
		\includegraphics[width=\linewidth]{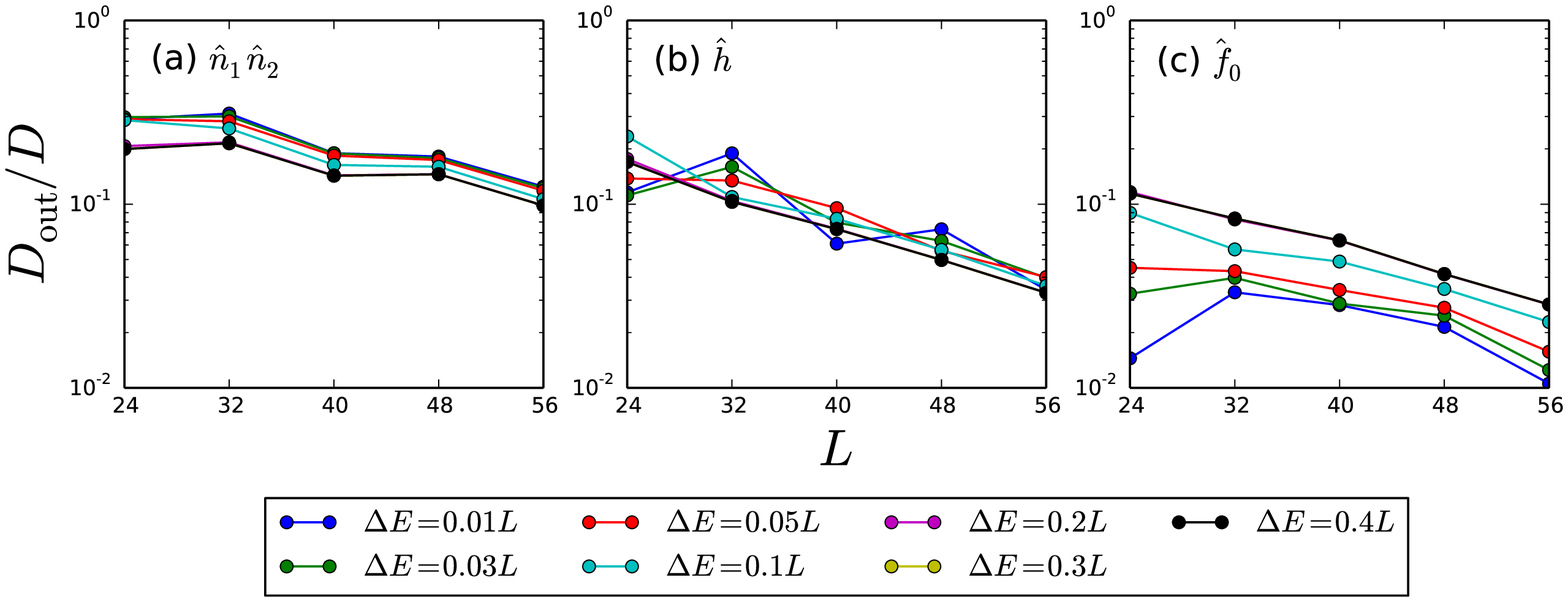}
		\caption{The $\Delta E$-dependence of $\dout/D$ for the integrable system.}
		\label{int_DeltaE}
	\end{figure}

	Figure \ref{int_DeltaE} shows the $\Delta E$-dependence of $\dout/D$. $\dout/D$ decays exponentially except for the case that $\Delta E$ is too small. When $\Delta E$ is too small, we cannot obtain the enough number of eigenstates for statistical analysis.
	
	In summary, the exponential decay of $\dout/D$ is universal and independent of the parameters $\varepsilon$, $E$, $\delta$ and $\Delta E$, except for some extreme cases.
	

	\newpage
	\subsection{Non-integrable system}
	We next show the parameter-dependence of $\dout/D$ for the non-integrable system (7) of the main text with $\lambda = 1$. In this model, the parameters used in the main text are given by $E=0$, $\Delta E=0.1L$, $\delta=0.1$, and $\varepsilon=0.003, 0.015, 0.05$ for $\hat{n}_1\hat{n}_2$, $\hat{h}$ and $\hat{f}_0$, respectively.
	
\begin{figure}[H]
		\centering
		\includegraphics[width=\linewidth]{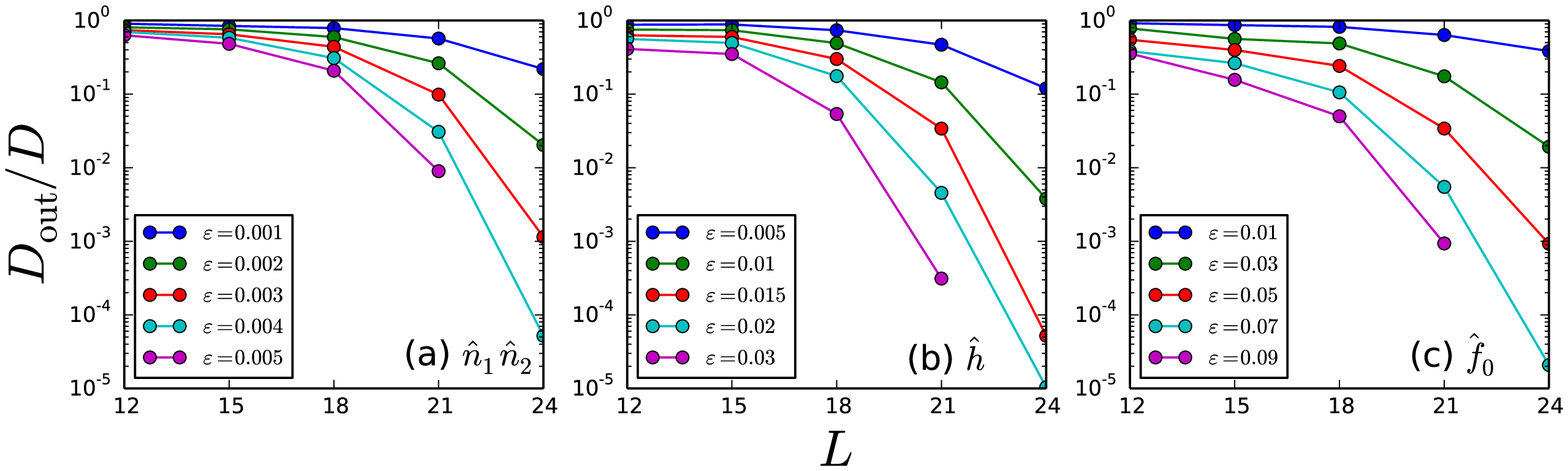}
		\caption{The $\varepsilon$-dependence of $\dout/D$ for the non-integrable system.}
		\label{non_epsilon}
	\end{figure}

	Figure \ref{non_epsilon} shows the $\varepsilon$-dependence of $\dout/D$. We see that $\dout/D$ decays double exponentially in $L$ regardless of $\varepsilon$. In contrast to the integrable system, $\dout/D$ with large $\varepsilon$ and large $L$ becomes exactly zero for all the observables. This is a clear evidence of the validity of the strong ETH. 
	
\begin{figure}[H]
		\centering
		\includegraphics[width=\linewidth]{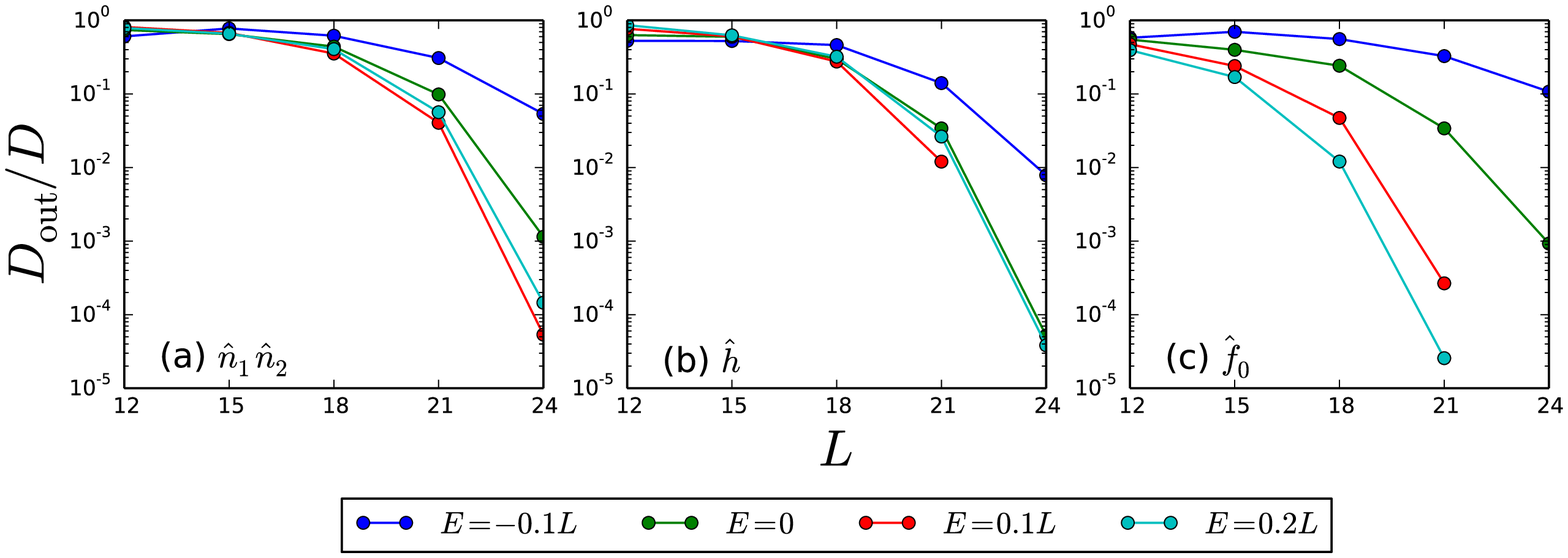}
		\caption{The $E$-dependence of $\dout/D$ for the non-integrable system.}
		\label{non_energy}
	\end{figure}

	Figure \ref{non_energy} shows the $E$-dependence of $\dout/D$. The double-exponential decay is seen in all the energy range.  
	
\begin{figure}[H]
		\centering
		\includegraphics[width=\linewidth]{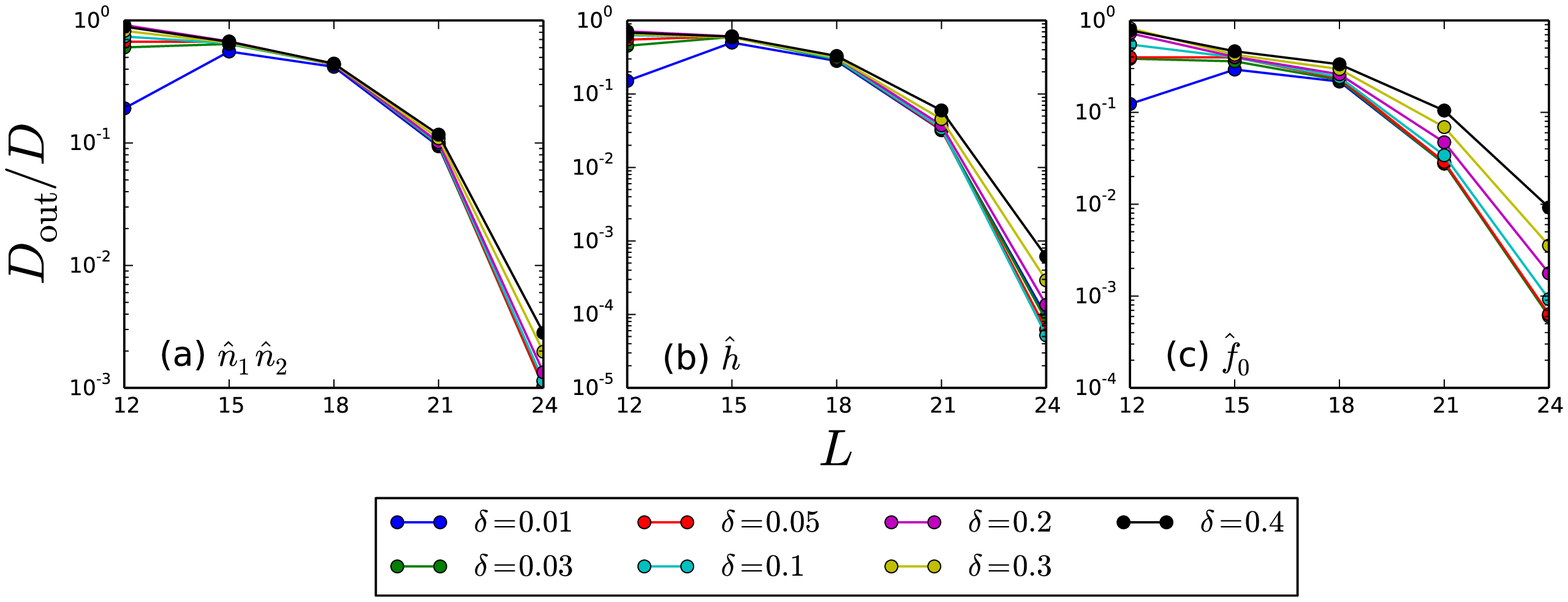}
		\caption{The $\delta$-dependence of $\dout/D$ for the non-integrable system.}
		\label{non_delta}
	\end{figure}

	Figure \ref{non_delta} shows the $\delta$-dependence of $\dout/D$. We found a similar trend to the results for the integrable system. If $\delta$ is too small, like $\delta=0.01$, $\dout/D$ does not show the double exponential decay for small $L$, because the number of eigenstates in the energy shell is too small. However, even with such a small $\delta$, $\dout/D$ decays double exponentially for large $L$.
	
\begin{figure}[H]
		\centering
		\includegraphics[width=\linewidth]{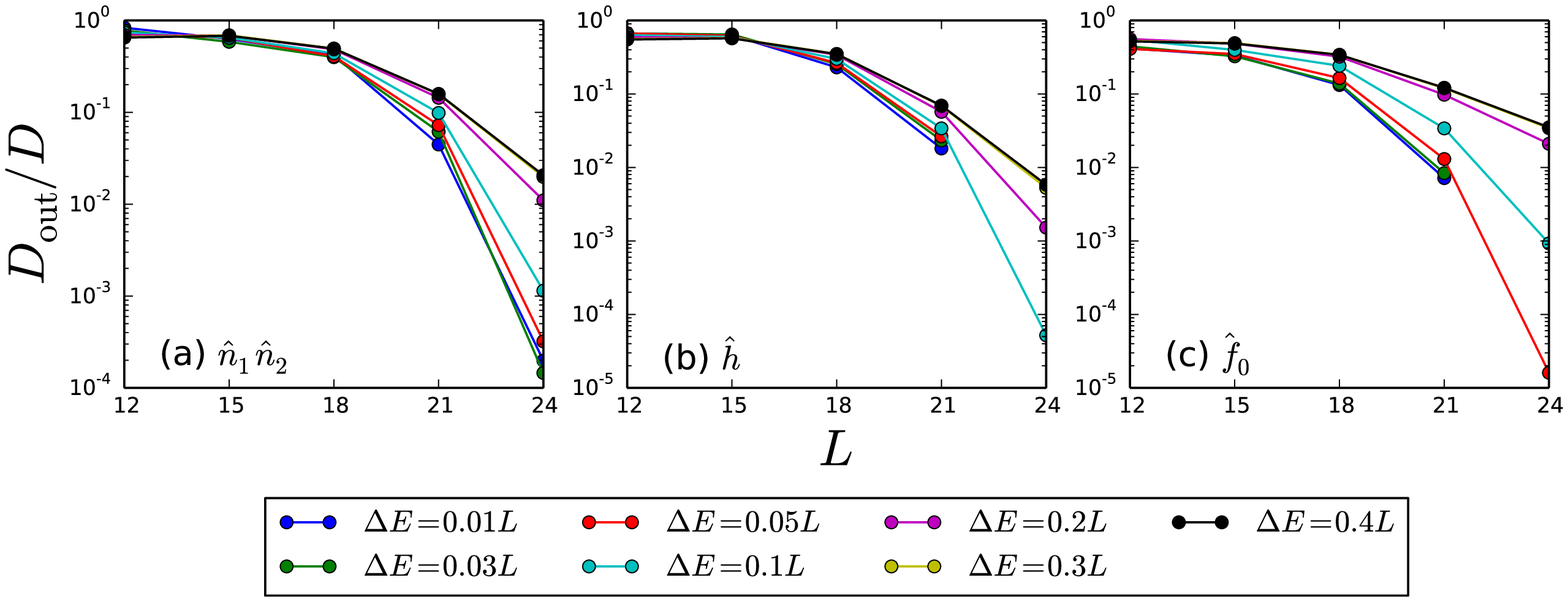}
		\caption{The $\Delta E$-dependence of $\dout/D$ for the non-integrable system.}
		\label{non_DeltaE}
	\end{figure}

	Figure \ref{non_DeltaE} shows the $\Delta E$-dependence of $\dout/D$. We see that $\dout/D$ decays double exponentially regardless of the value of $\Delta E$.
	
	In the above results, the double exponential decay is universally seen for all the parameters except for a few extreme cases, which supports our conclusion that the strong ETH holds for non-integrable systems independently of the choices of the parameters.
	
	
	\section{The operator norm of the commutator}
\begin{figure}[H]
		\centering
		\includegraphics[width=0.5\linewidth]{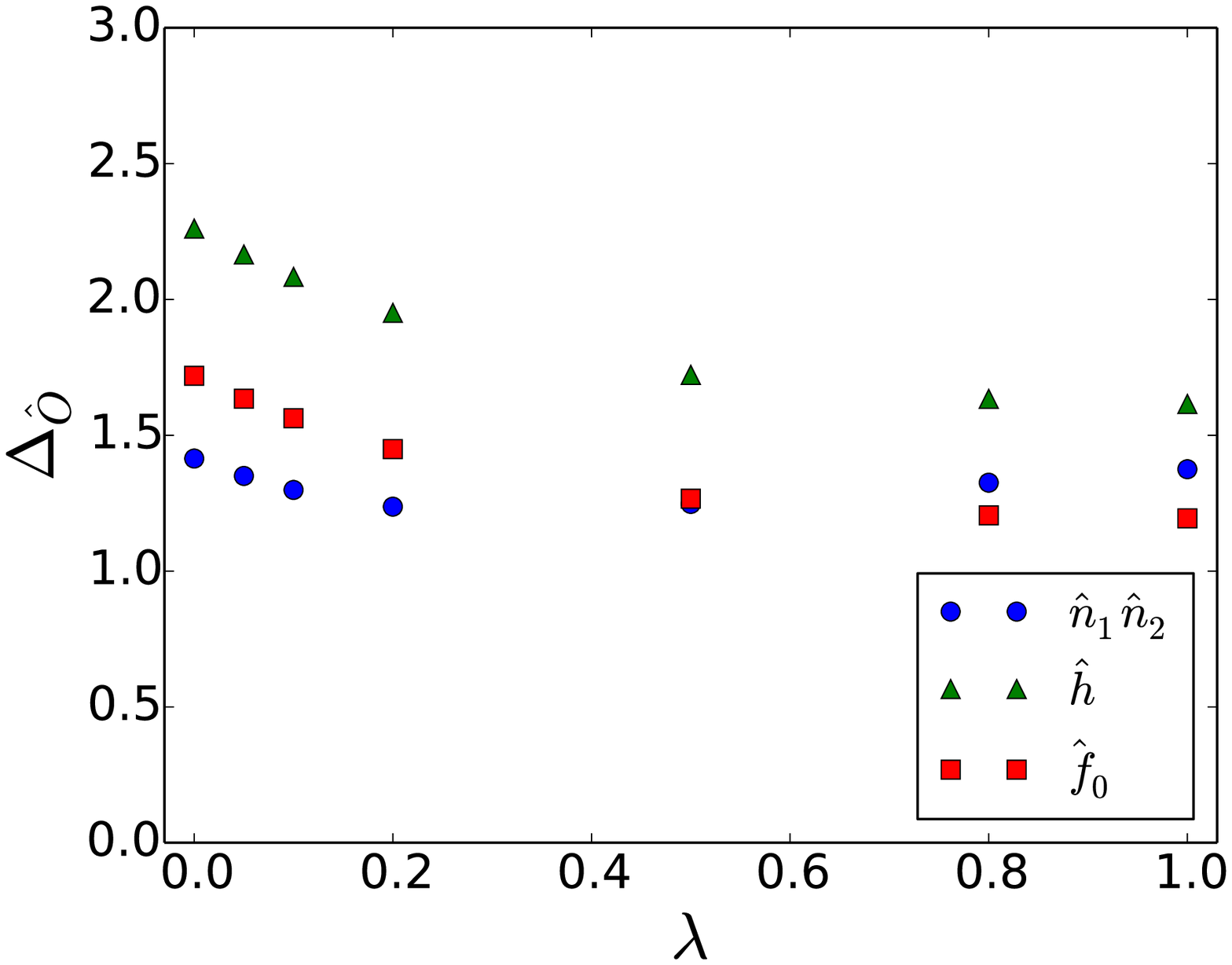}
		\caption{The $\lambda$-dependence of $\Delta_{\hat O}$, which is numerically obtaind for $L = 21$.}
		\label{norm}
	\end{figure}

	In this section, we show the operator norm of the commutator $\Delta_{\hat O} := \|i[\hami_{XXX}, \hat O]\|/\|\hat O\|$ for $\hat O = \hat n_1\hat n_2$, $\hat h$, and $\hat f_0$. The norm $\Delta_{\hat O}$ quantifies a ``distance'' between the observable $\hat O$ and the Hamiltonian $\hami_{XXX}$, and therefore a smaller value of $\Delta_{\hat O}$ implies smaller effective randomness of $\hat O$ with respect to $\hami_{XXX}$. Figure~\ref{norm} shows the $\lambda$-dependence of $\Delta_{\hat O}$. We find that for $\lambda \geq 0.5$, the largest value of $\Delta_{\hat O}$ is given by $\hat h$, followed by $\hat n_1\hat n_2$ and $\hat f_0$ in descending order. This order is the same as that of $c$ and $sz$ in Fig~4~(b) and (c) in the main text. This result supports our argument that a smaller distance between $\hat O$ and $\hami_{XXX}$ implies smaller values of $c$ and $sz$. Furthermore, for $\lambda \geq 0.5$, $\Delta_{\hat O}$ is almost independent of $\lambda$. This is consistent with Fig.~4~(b) in the main text, in which $c$ is also almost independent of $\lambda$ in the same range.
	
	\section{The $\lambda$-dependence of $s$}
	We here show the values of entropy density $s$ in detail. We numerically evaluated $s$ through the fitting of $D:= |\mathcal{M}(0, 0.1L)|$ by $f(L) = ae^{sL}$. Table~\ref{entropy_density} shows the results of the fitting, where $s$ approximately equals $0.6$ and is almost independent of $\lambda$. 
	
	\begin{table}[H]
		\centering
		\begin{tabular}{|c||c|c|}
			\hline
			$\lambda$& $s$ & Asymptotic Standard Error\\
			\hline\hline
			1 &	0.598202  &	0.0008833\\
			0.8 & 0.597839 &	0.0002985\\
			0.5 &	0.601356 &	0.0005085\\
			0.2&0.608291&0.0006695 \\	0.1&0.609596&0.0009544 \\	0.05&0.616175&0.002512\\
			0&0.615605&0.001714\\
			\hline
		\end{tabular}
		\caption{The $\lambda$-dependence of $s$.}
		\label{entropy_density}
	\end{table}
	
\end{document}